\documentclass[preprint,amsmath,amssymb]{revtex4}
\usepackage{graphicx}
\usepackage{color}
\usepackage{bm}

\newcommand{\cyb}{{C${}_6$Yb}}
\newcommand{\cca}{{C${}_6$Ca}}

\def\<{\langle}
\def\>{\rangle}
\makeatother
\begin{document}

\title{Electronic structure of superconducting graphite intercalate
  compounds: The role of the interlayer state}

\author{G\'abor Cs\'{a}nyi, P. B. Littlewood, Andriy H. Nevidomskyy, C. J. 
Pickard and B. D. Simons} 
\affiliation{Cavendish Laboratory, Madingley Road, Cambridge CB3\ OHE, UK}

\date{\today}

\maketitle 

{\bf Although not an intrinsic superconductor, it has been long--known
that, when intercalated with certain dopants, graphite is capable of
exhibiting superconductivity~\cite{Dresselhaus_review}. Of the family
of graphite--based materials which are known to superconduct, perhaps
the most well--studied are the alkali metal--graphite intercalation
compounds (GIC)~\cite{Koike} and, of these, the most easily fabricated
is the C${}_8$K system~\cite{Hannay} which exhibits a transition
temperature $\bm{T_c\simeq 0.14}\,$K~\cite{Koike}. By increasing the alkali
metal concentration (through high pressure fabrication techniques),
the transition temperature has been shown to increase to as much as
$\bm 5\,$K in C${}_2$Na~\cite{Belash}. Lately, in an important recent
development, Weller \emph{et al.} have shown that, at ambient conditions,
the intercalated compounds \cyb~and \cca~exhibit superconductivity with 
transition temperatures $\bm{T_c\simeq 6.5}\,$K and $\bm{ 11.5}\,$K 
respectively~\cite{Weller}, in excess of that presently reported for other
graphite--based compounds. We explore the architecture of the states
near the Fermi level and identify characteristics of the electronic
band structure generic to GICs. As expected, we find that charge transfer 
from the intercalant atoms to the graphene sheets results in the occupation
of the $\bm\pi$--bands. Yet, remarkably, in all those -- and only those -- 
compounds that superconduct, we find that an interlayer state, which is 
well separated from the carbon sheets, also becomes occupied.
We show that the energy of the interlayer band is controlled by a
combination of its occupancy and the separation between the carbon
layers.}

\begin{figure}
\includegraphics[width=6.5in]{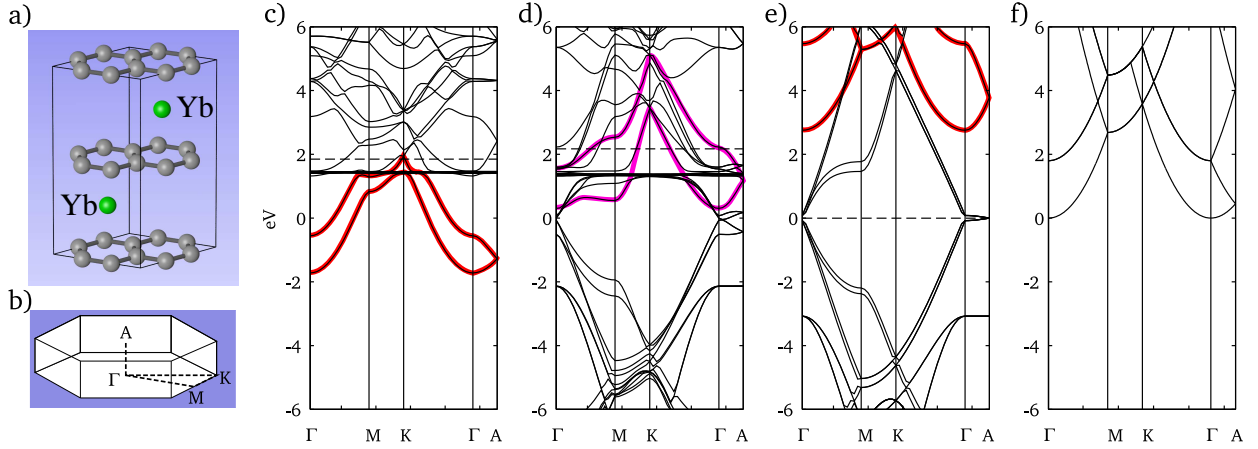}
\caption{Band structure of \cyb~compared to various model systems as
inferred from DFT--LDA. The panels correspond to (a) atomic structure
and (b) geometry of the Brillouin zone of \cyb, band structures of (c)
expanded Yb, with the lattice parameters and atomic positions
corresponding to those in \cyb, so that the nearest Yb-Yb distance is
increased by 24\% as compared to the pure FCC structure, 
(d) the \cyb~compound, (e) the
``empty'' graphite system, where the Yb atoms have been removed but
the lattice constants are kept fixed, and (f) free electron bands,
folded into the unit cell of \cyb.  The free electron-like band is
marked with a thick red line, while pink lines indicate hybridized
bands with significant charge density in the interlayer region.  The
band structures (except for the free electron bands) are aligned with
respect to the core levels and in each case the Fermi level is shown
with a dashed line.  Including correlation effects for the localized
$4f$ orbitals within the LDA+U method results only in the expected
downward rigid shift of the energy of the filled $4f$ levels (not
shown).}

\label{Fig:mainbs}
\end{figure}

To focus our discussion, we begin with an investigation of the 
newly discovered \cyb\ system. Our calculations rely on density functional 
theory (DFT) techniques~\cite{dft} applied within the local density 
approximation (LDA)~\cite{lda}. Since the $4f$-shell of Yb is found to 
be filled (in agreement with experiment), the DFT--LDA can be applied 
with some confidence. Moreover, one may infer that the $4f$-orbitals 
play no essential role in superconductivity. The atomic 
structure of \cyb, shown in Figure~\ref{Fig:mainbs}a, involves 
a stacked arrangement of graphene sheets intercalated with a triangular 
lattice configuration of Yb atoms. The stacking of the graphene sheets 
is AAA, while the Yb atoms occupy interlayer sites above the centres 
of the hexagons of the graphene sheets in an ABAB stacking arrangement 
leading to a P63/MMC crystallographic structure. We use the same geometry 
for \cca.

\begin{figure}
\includegraphics[width=6.5in]{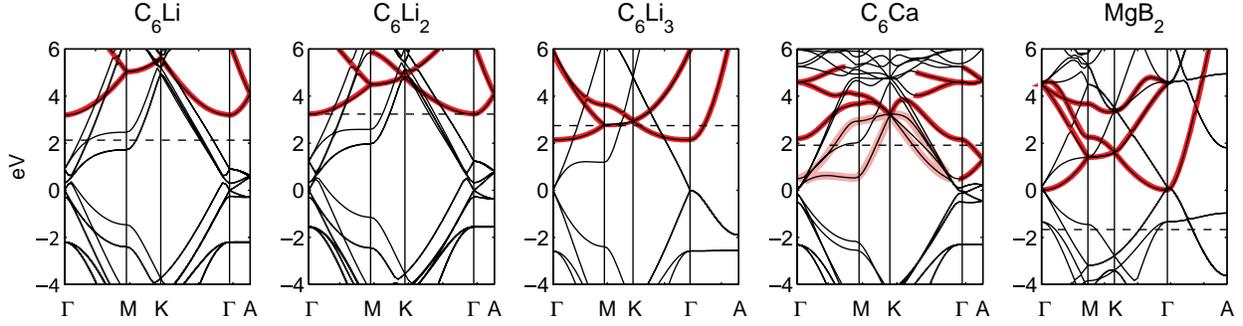}
\caption{Band structures of the Li series of GICs, \cca~and
  MgB$_2$. The latter is shown in a supercell that matches the
  others in the plane of the graphene sheets.  The interlayer band is
  marked with a thick red line, while pink lines indicate hybridized
  bands with significant charge density in the interlayer region.  The
  band structures are aligned by the crossing of the $\pi$ bands, in
  each case the Fermi level is shown with a dashed line.}
\label{Fig:bscompare}
\end{figure}

Figure~\ref{Fig:mainbs}d shows the results of the band structure 
calculation of \cyb~centered on the Fermi energy. To resolve the 
qualitative structure of the levels and identify the bands intersecting 
the Fermi level, it is helpful to compare the dispersion with the 
corresponding ``empty'' graphite--like system and the ``empty'' Yb 
metal, obtained by simply removing, respectively, the Yb and carbon 
atoms from the \cyb\ structure whilst holding the lattice constants 
fixed. 
In \cyb, the previously 
unoccupied graphite $\pi^*$ bands 
are seen to intersect the new Fermi level (due to the enlarged interlayer 
distance each are almost doubly degenerate). Of these, one remains almost 
unperturbed by the presence of Yb while the degeneracy of the other is 
lifted, and one of the resulting bands strongly hybridizes with a new 
band that also crosses the Fermi level. 
The latter resembles both the free electron-like state of expanded Yb
metal (panel(c)) and the interlayer band intrinsic to
pure graphite (panel (e)).
As well 
as the $5s$ and $5p$~orbitals (which lie well outside the range of 
energies shown in Figure~\ref{Fig:mainbs}d), the $2\times 7$ $4f$~orbitals 
associated with the two Yb atoms in the unit cell form a set of 
localized almost non--dispersive bands which appear at ca.~$0.8$~eV 
below the Fermi level. Referring to Figure~\ref{Fig:bscompare}~(panel 4), 
and the discussion in Ref.~\cite{CalandraMauri}, we note that a
similar phenomenology applies to the band structure of \cca.

\begin{figure}
\bigskip
\includegraphics[width=6.5in]{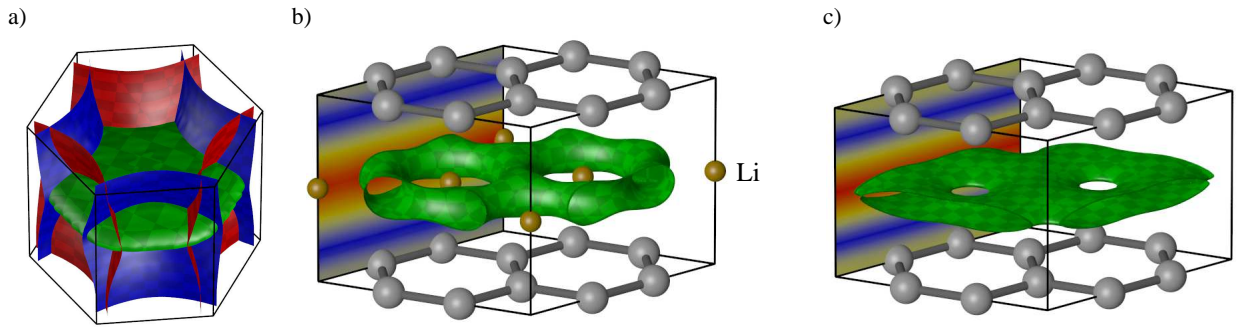}
\caption{Structure of the interlayer state. (a) Fermi surface of the
  C$_2$Li system as inferred from the band structure calculation. 
  In its native state, the Fermi surface of a single graphene sheet
  collapses onto two points in the Brillouin zone. Electron-doped, the
  Fermi surface expands to form a set of two cylindrical electron--like
  surfaces associated with the $\pi^*$ bands which enclose the singular 
  points in the spectrum (red and blue). By contrast, the interlayer 
  band, which adopts a nearly free--electron like structure parallel 
  to the planes near the $\Gamma$ point, acquires a more complex extended 
  character at the Fermi surface (green).
  (b) An isosurface of the charge density associated with the lowest
  interlayer band in C$_2$Li at the $\Gamma$ point. The colourmap on
  the left plane shows the projection of the band density, with blue
  corresponding the low and red to high electron density. The spatial
  structure of the interlayer band is essentially identical to that of
  its unoccupied analogue in pure graphite, which is shown in panel (c).}
\label{Fig:fermis}
\end{figure}

Although the presence of the interlayer band and its hybridisation 
with the $\pi^*$ band helps to explain the diminished normal state 
resistance anisotropy observed in \cyb\ over that of pure 
graphite~\cite{Weller}, could its occupation be significant for 
superconductivity? To address this question, it is instructive to 
draw comparison with other superconducting GICs. It is known from LEELS 
measurements of the superconducting compound C${}_8$K that, in common 
with the \cyb~system, the interlayer band lies below the Fermi 
energy~\cite{Koma}. However, a more discerning test of the significance 
of the interlayer state is presented by the Li intercalates. While 
intercalation at ambient pressure leads to the C$_6$Li structure, high 
pressure techniques can be used to fabricate C$_3$Li and C$_2$Li compounds. 
Significantly, while the C$_3$Li system remains (like its cousin 
C$_6$Li) non-superconducting, C$_2$Li exhibits a transition with a 
$T_c\simeq 1.9$\,K~\cite{Belash2}. Referring to
Figures~\ref{Fig:bscompare} and \ref{Fig:fermis}, 
one may note that, in common with the \cyb~and \cca~system, the Li 
GICs also exhibit a free electron-like interlayer band. 
The dispersive character of this band is reflected clearly 
in the Fermi surface (see Fig.~\ref{Fig:fermis}a).
However, significantly, increasing Li ion concentration (and, with 
it, the degree of electron doping) leads to a lowering of the interlayer 
band resulting in its occupancy in the C$_2$Li system. 
Indeed, combining the available data in Table~\ref{Tab:GIC}, the coincidence of 
superconductivity in the GICs with the occupation of the interlayer 
state is striking.

\begin{table}
\begin{ruledtabular}
\begin{tabular}{lrrrr}
GIC& electron & interlayer & interlayer & $T_c$ \\
   & doping   & separation & band occ.  &       \\
\hline
graphite   & 0   &$3.35$\AA  & no   & ---   \\   
C$_6$Li & 1/6 & $3.7$\AA     & no   & ---   \\
C$_3$Li & 1/3 & $3.7$\AA     & no   & ---   \\
C$_2$Li & 1/2 & $3.7$\AA    & yes  & $1.9$K\\
C$_{16}$K  & 1/16 & $5.2$\AA     & no  & ---\\
C$_8$K  & 1/8 & $5.2$\AA     & yes  & $0.14$K\\
C$_6$Ca & 1/3 & $4.6$\AA     & yes  & $11.5$K\\
C$_6$Yb & 1/3 & $4.7$\AA     & yes    & $6.5$K\\
C$_6$Ba & 1/3 & $5.25$\AA    & yes  & ?\\
\end{tabular}
\end{ruledtabular}
\caption{Table recording the coincidence of the interlayer state occupation 
in GICs and the phenomenon of superconductivity as inferred from the results
of band structure calculation and experiment. According to the electronic
structure calculations and the observed phenomenology, one would predict 
superconductivity in the GIC C$_6$Ba~\cite{Herold}.}
\label{Tab:GIC}
\end{table}

Studies of the free electron-like interlayer state in pure graphite
and GICs have a long history. Although present in early band structure
calculations in the pure system, the significance of the (unoccupied)
band in graphite and C${}_6$Li was first emphasized by Posternak
\emph{et al.}~\cite{Posternak,Holzwarth84} while, subsequently, its
existence was confirmed experimentally in both graphite~\cite{Reihl}
and C${}_6$Li~\cite{Fauster}. (The same band has been equivalently referred 
to as a metallic ``sd band'' in Ref.~\cite{Molodtsov}.) Referring to the 
Li based GICs, it is apparent that the energy, and therefore also the
occupancy, of this band can vary considerably with respect to the other 
bands present in the above systems. 

\begin{figure}
\includegraphics[width=5.0in]{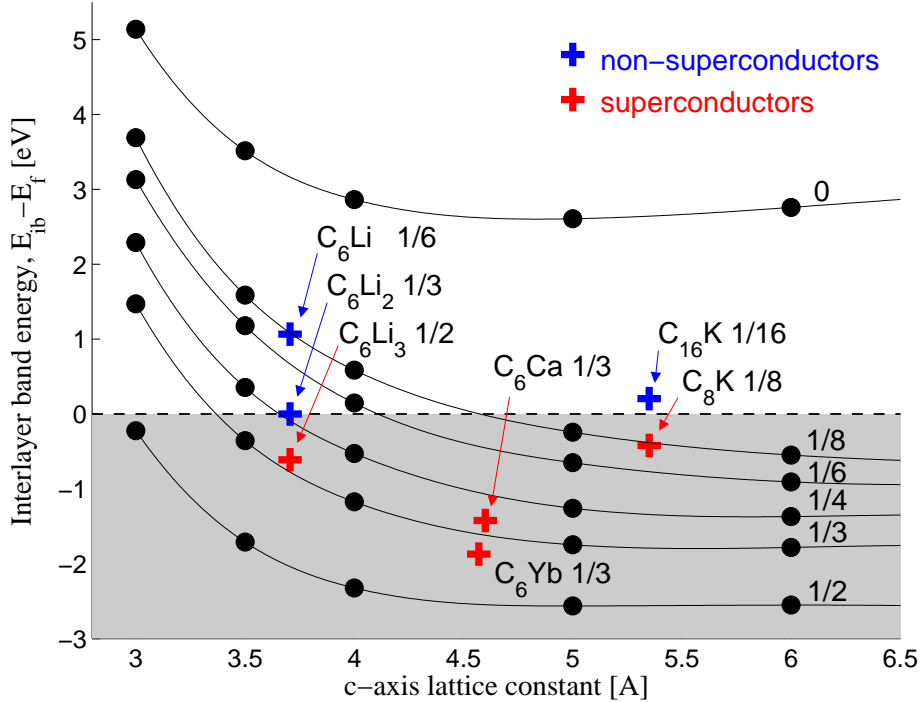}
\caption{The dependence of the interlayer band energy (at the $\Gamma$
 point) of the empty graphite system with changing $c$ axis spacing
 for different electron dopings. (Note that the calculation of the
 electron doped system using periodic boundary conditions requires
 overall charge neutrality, enforced by a uniform positive
 background charge.)  Superposed on the figure are data points corresponding
 to different GICs, red and blue crosses indicate whether the compound
 is found to be superconducting or not, respectively. The fractions
 show the electron doping of the graphite sheets in units of electron
 per carbon atom. In the systems that fall into the shaded area, the
 interlayer band is occupied.}
\label{Fig:dop_sep}
\end{figure}

It is notable that the dispersive 5$d$ band of Yb in the expanded metal 
(Fig.~\ref{Fig:mainbs}c) and the interlayer state in pure graphite 
(Fig.~\ref{Fig:mainbs}e) both bear strong resemblance to the dispersion 
of the band of free electrons in the absence of external potential 
(Fig.~\ref{Fig:mainbs}f). 
Indeed, we can understand 
the important features of this nearly free-electron like state by 
studying the ``empty'' graphite system with the intercalant ions removed:
In Figure~\ref{Fig:dop_sep}, we show the effect of charging and layer 
separation on the energy of the resulting interlayer band. An increase 
in the $c$ axis lattice constant and in the accumulated charge both 
lead to a lowering of the band energy, and thus an increase in its 
occupancy. It is interesting to note that, for a wide range of 
experimentally known GICs, the interlayer band energy follows closely 
that of the pure graphite model.

Motivated by preliminary electronic structure calculations, previous
theoretical studies by Al Jishi~\cite{Jishi83,Jishi92} have emphasized 
the significance of partially occupied two--dimensional graphite 
$\pi^*$ bands and three--dimensional interlayer bands for 
superconductivity. Indeed, these studies have shown that aspects of 
superconductivity in some GICs can be described well by a two--band 
phenomenology. Yet the underlying microscopic pairing mechanism remains 
in question. Significantly, unlike the fullerene-based superconducting 
compounds, the $\pi^*$ bands are decoupled from the out-of-plane lattice 
vibrations of the graphene sheets, and couple only weakly to the 
in-plane modes. By contrast, one may expect the interlayer band to 
engage with lattice vibrations of the metal ions and, through 
their hybridisation with the $\pi^*$ bands, the graphene sheets. 
Indeed, recent numerical calculations by Calandra and Mauri have 
revealed the presence of strong electron--phonon coupling in \cca, 
facilitated by the occupation of the free-electron like 
band~\cite{CalandraMauri}. Whether an electron--phonon mechanism alone 
can explain the broad distribution of $T_c$ observed across the range 
of GICs, as well as the staging dependence~\cite{Iye}, remains a 
question for future investigation. In this context, it is interesting 
to note that, in those compounds with a low transition temperature, the 
interlayer band remains nearly orthogonal to the $\pi^*$ bands, as it 
is in pure graphite, while in \cyb~and \cca~the hybridization is strong. 

Intriguingly, through its weak coupling with the graphene layer, one may 
expect occupancy of the intrinsic interlayer band of graphite to provide
an ideal environment for soft charge fluctuations that could promote 
($s$--wave) superconductivity via an excitonic pairing 
mechanism~\cite{Bardeen}. Indeed, the same weak coupling diminishes the 
potential for charge density wave instability which might otherwise 
compete with the superconducting pair correlations. In this context, 
it is interesting to note that a parallel pairing mechanism involving 
the exchange of acoustic plasmons was suggested by Takada~\cite{Takada} 
and was much discussed after the discovery of superconductivity in 
the cuprates. Recently, the model has been re-evaluated for layered 
metals~\cite{Bill03}.
Bearing in mind that low-frequency plasma 
oscillations will, of course, couple to the ionic positions---conventional 
acoustic phonons can be viewed as simply coupled ionic and electronic 
plasma oscillations---such processes present an effective mechanism to 
enhance electron--phonon coupling. 

Recognizing the auspicious role played by the free-electron like state in the
superconducting GICs, it is tempting to look for a common phenomenology 
in the isoelectronic family of boron intercalcates and, in particular, 
the well--studied superconducting compound MgB${}_2$ which exhibits a 
transition temperature $T_c=39$K~\cite{Nagamatsu}. Although the material 
indeed possesses interlayer states, the band structure 
(Fig.~\ref{Fig:bscompare}) shows that it remains unoccupied. However,
in contrast to \cyb~and the other superconducting GICs, in this case
charge carriers are transferred from the $\sigma$ to the $\pi^*$
band. The resulting band of hole--like $\sigma$ states couples strongly
to the host lattice providing a natural environment to realise a high
temperature phonon--mediated superconductivity.

Finally, it is interesting to note that the interlayer states find
their analogue as nearly free--electron like states in the carbon
nanotube system: early theoretical~\cite{Miyamoto} and
experimental~\cite{Shimoda} investigations of the alkali--metal
nanotube intercalates already indicate a phenomenology in doping
dependence similar to the GICs. On this background, an investigation
of possible superconductivity in electron--doped carbon nanotubes would
seem to be a timely and worthwhile enterprise.

\underline{Methods:} The results reported here were obtained from an
implementation of the LDA within the framework of the Castep 
package~\cite{castep}. The methodology, which includes the use of
ultrasoft pseudopotentials, has been validated in many environments 
including the rare earth compounds~\cite{RareEarthsCastep}. Here, we 
take advantage of the fact that, in using ultrasoft pseudopotentials, 
one can exercise a considerable degree of freedom in deciding which 
electron shells to treat as fixed core and which to treat as valence, 
free to participate in bonding.

The numerical results presented for \cyb~were computed treating all
the electrons below the $5s$ level for Yb, and below the $2s$ level
for carbon atoms as core, respectively.  Plane waves up to $480\,$eV
were used to expand the electronic states, and the Brillouin zone was
sampled with a $12\times 12\times 6$ mesh. (A test calculation using a
$600\,$eV cutoff did not yield appreciably different results.) To
check for possible correlation effects, we used the LDA+U method as
implemented in the PWSCF package~\cite{PWSCF} with $U$ varying between 
3 and 7 eV.

{\sc Acknowledgements:} We are grateful to S. S. Saxena and Mark Ellerby
for bringing the \cyb~and \cca~compounds to our attention and to Gil 
Lonzarich for contributing to our understanding. 

\end{document}